# A New Approach in Optimal Control of Step-Down Converters Based on a Switch-State Controller

Ali Parsa Sirat
*Department of Electrical and Computer Engineering*
University of North Carolina at Charlotte
Charlotte, North Carolina
aparsasi@uncc.edu

Niloofar Zendehdel
*Department of Electrical and Computer Engineering*
Babol Noshirvani University of Technology
Babol, Iran
niloofar.zendehdel@gmail.com

*Abstract*— **In this paper, an optimal approach based on on-off controller is used to optimally control a DC-DC step-down converter. It is shown that the conventional controller techniques of DC-DC converters based on a linearized averaging model have several drawbacks, including different operating mode, linearization concerns, and constraint difficulties. A single-mode discretize state-space model is used, and a new optimal approach policy is implemented to control a step-down DC-DC converter. The simulation results confirm that the proposed DC-DC model associated with the optimal controller functions properly in controlling a DC-DC unfixed switch-mode step-down converter that is facing load changes, noisy inputs, and start-up procedure.**

*Keywords—switching, control, style, buck, power electronics converter (key words)*

## I. Introduction

Switch-mode DC-DC converters are widely used in a vast variety of industries and applications, such as photovoltaics [1], [2] and battery energy storages [3], as a mature and well-established energy conversion technology. These converters are widely discussed in power systems, renewable energies, control theory, and transformer designs [4], [5]. However, several control concerns and challenges are still presented in a great number of publications and research papers each year. The improvement in computational power and the enhancement in power control field of study have encouraged researchers to tackle the concept from new perspectives.

The hybrid nature of DC-DC converters is the main root of DC-DC converter concerns. The step-down DC-DC converter has three operation modes, and each mode has its own linear dynamics. Moreover, there are different constraints that make the control design more complicated. Natural constraints, for example the duty cycle, which is bounded between zero and one, or the nonnegative inductor current limitation. In addition, there are some safety-related constraints such as current limitation or soft-switching. Lastly, the variation in the input and the output voltage makes the control designing process more complicated since it changes the operating point .

The state-space averaging model is the most common approach to modeling the DC-DC converters, and proportional-integral (PI) type with the pulse width modulation (PWM) unit is the dominant controller [1], [2]. The controller is designed for the linearized model around the system's operating point. In the literatures, a great variety of approaches has been studied for improving the controller design; nevertheless, the averaging model is the base of most of these approaches. In this category, the methods introduced vary from linear quadratic regulators (LQR) [3] to fuzzy logic [4]. Several feedforward control [5], [6], and nonlinear control [7]-[10], and neuro-control techniques [11]-[13] have been studied in recent years. The author in [14] proposed a nonlinear predictive controller for an unconstrained DC-DC power converter. An LQR approach is proposed in [14], [15] and [29]-[36], which uses the discrete-time linearized model of the average DC-DC converter. This method of control can consider various sorts of power loss and battery storage energy management [16]-[28]. More critical challenges is recently analyzed in Smart and Microgrids [36]-[44].

To overcome the aforementioned concerns, this paper presents a new optimal approach in modelling and control designing for the DC-DC converter. The proposed optimal approach eliminates the concerns of having three different modes, eliminates the linearization model around operating points, and it tackles a constrained DC-DC converter.

The rest of this paper is organized as follows. First, the problem statement is presented in Section II. The optimal control concept, including the Bellman equation and its solution, is explained in Section III. The optima controller is simulated and evaluated in Section IV. Finally, the conclusion is presented in Section V.

## II. Problem Statement

Consider a stable linear time invariant system represented by the discrete-time state-space equation

$$x_{k+1} = Ax_k + bu_k, \qquad k = 0,1,2,\ldots \qquad (1)$$

with the state vector $x_k \in \mathbb{R}^n$ and the scalar control input $u_k$. Here, $A$ is a constant $n \times n$ matrix, $b$ is a constant $n \times 1$ vector, and $(A, b)$ is assumed controllable. The system is driven by an on-off actuator that constrains the control input $u_k$ to take values in the binary control set $\mathcal{U} = \{0,1\}$. For the purpose of state feedback, the state vector is measured by an array of precise sensors.

The objective is to develop a feedback law to regulate the state of the system around a desired constant set point $r \in \mathbb{R}^n$ (i.e., to maintain the error vector $e_k = x_k - r$ is reasonably small in some reasonable sense). This goal must be achieved with a reasonably small number of the actuator switching (transition from on to off or vice versa). This latter requirement is posed by the actuator structural limitations or is deliberately imposed by a designer to reduce energy losses in actuators such as solid-state electric switches.

These objectives (in general opposition) can be incorporated into a performance measure represented by an infinite horizon cost function. Then, design of an appropriate control law is formulated as a standard optimal control problem seeking the minimum of this cost function, as stated in the remainder of this section. By resolving this standard problem in Section III, an optimal control law is developed.

In this paper, the distance between the state vector $x_k$ and the set point $r$ is measured by the quadratic form

$$q(x_k) = (x_k - r)^T Q(x_k - r), \qquad (2)$$

where $Q$ is an $n \times n$ positive semidefinite matrix. A special form of this measure can be used to formulate an output regulation problem. Let $h$ be a $1 \times n$ vector and consider the scalar output:

$$y_k = h x_k, \quad k = 0,1,2, \dots. \qquad (3)$$

The function of an output regulator is to maintain this output near a fixed set point $s \in \mathbb{R}$. The distance between $y_k$ and $s$ is measured by $\|h x_k - s\|^2$, which can be expressed as (1) with the positive semidefinite $Q = h^T h$ and $r = h^T(h h^T)^{-1} s$.

In order to count the number of control transitions, a binary variable $z_k$ is introduced to keep track of the actuator state at the preceding time step. This new state variable evolves in time according to the state-space equation:

$$z_{k+1} = u_k, \quad k = 0,1,2, \dots \qquad (4)$$

$$z_0 = 0. \qquad (5)$$

In terms of $u_k$ and $z_k$, the number of control transitions in the time step $k$ is given by $|u_k - z_k|$.

The running cost (cost per stage) is defined as a function of the extended state $(x_k, z_k)$ and the control $u_k$ as

$$c(x_k, z_k, u_k) = (x_k - r)^T Q(x_k - r) + \beta |u_k - z_k|, \qquad (6)$$

where _ is a positive constant. This function simultaneously penalizes both the deviations of state from their set point and the number of control transitions. The relative importance of these extremes can be adjusted by the design parameter $\beta$.

The control performance is measured in this paper by the discounted infinite horizon cost function:

$$J = \sum_{k=0}^{\infty} \alpha^k c(x_k, z_k, u_k), \qquad (7)$$

where $0 < \alpha < 1$ is a constant discount rate. By introduction of the discount rate $\alpha$, the relative importance of the running cost in the near future versus far future can be adjusted when making a control decision at the present time. An even more important issue is that the infinite sum (7) does not converge without a discount rate (i.e., for $\alpha = 1$), since the running cost in (7) generally does not converge to 0 as $k \to \infty$.

The optimal control problem considered in this paper is defined as follows: Determine a feedback control law of the form

$$u_k = \mu(x_k, z_k) \qquad (8)$$

to minimize the discounted infinite horizon cost function

$$J = \sum_{k=0}^{\infty} \alpha^k ((x_k - r)^T Q(x_k - r) + \beta |u_k - z_k|), \qquad (9)$$

subject to the state-space equations

$$x_{k+1} = A x_k + b u_k \qquad (10)$$

$$z_{k+1} = u_k \qquad (11)$$

with a given initial state $x_0$ and $z_0 = 0$.

III. OPTIMAL CONTROL

In the following section, we derive a policy to serve the proposed optimal problem formulation in Section II. In Section III-A, the Bellman equation and the corresponding optimal policy is presented, and in Section III-B, the solution to the proposed optimal law is demonstrated.

A. Bellman Equation

The Bellman equation can solve the optimal problem for a bounded cost per stage. The proposed cost per stage based on (10) is the sum of two parts, and therefore by showing that each part is bounded, the cost per stage is bounded.

The first part, $q(x_k)$ is bounded because the system is stable and the states and the reference are bounded. The second part, $\beta |u_k - z_k|$, is also bounded since $u_k, z_k \in \{0,1\}$ and $\beta$ is a bounded coefficient. Consequently, the proposed cost per stage is bounded and the Bellman equation can give the corresponding optimal policy. The Bellman equation for the proposed optimal policy can be written as

$$V(x, z) = \min_{u \in \{0,1\}} \{c(x, z, u) + \alpha V(Ax + bu, u)\}, \qquad (12)$$

where $V(x, z)$ is the value function. Furthermore, the optimal policy can be defined as

$$\Psi^*(x, z) = \underset{u \in \{0,1\}}{\operatorname{argmin}} \{c(x, z, u) + \alpha V(Ax + bu, u)\} \qquad (13)$$

where $\Psi^*(x, z)$ is the stationary optimal policy that is a function of the value function. For future simplification, a new notation is defined as

$$V_i(x) = V(x, i), \quad i \in \{0,1\} \qquad (14)$$

where $V_i(x)$ represents the value function when the second input is equal to $i$.

Taking into account that q(x) is not a function of u, it can be shifted out in (14), and to find the minimum term we can replace u with two possible choices. Applying these two, (14) can be written as

$$V(x, a) = q(x) + \min\{\beta z + \alpha V_0(Ax), \beta(1 - z) + \alpha V_1(Ax + b)\}. \tag{15}$$

In order to simplify the equation, another notation is defined as

$$f(x) = V_1(Ax + b) - V_{0(Ax)}. \tag{16}$$

To find the minimum argument if the first term is less than the latter one, $u = 0$ should be chosen; otherwise, $u = 1$ is the optimal solution. Hence by comparing the two values in the minimizing part in (15) and using the defined notations, the optimal policy can be rewritten as

$$\Psi^*(x, z) = \begin{cases} 1, & f(x) \leq \frac{1}{\alpha}\beta(2z - 1) \\ 0, & f(x) > \frac{1}{\alpha}\beta(2z - 1). \end{cases} \tag{17}$$

As shown in (17), the optimal policy, $\Psi^*(x, z)$, is dependent on $f(x)$. In the other words, it is dependent on the value function. Therefore, the value function needs to be calculated by solving (15).

### B. Solution to the Bellman Equation

In this part, the optimal solution corresponding to (15) is proposed. To simplify the equation we are using the fact that $z$, as the last state of control, is constrained in the set of $\{0,1\}$, so by replacing the values of z, the Bellman equation can be written as two separate equations as

$$V_0(x) = q(x) + \min\{\alpha V_0(Ax), \beta + \alpha V_1(Ax + b)\}, \tag{18}$$

$$V_1(x) = q(x) + \min\{\beta + \alpha V_0(Ax), \alpha V_1(Ax + b)\} \tag{19}$$

Solving this equation analytically is difficult, but if it can be changed to a symmetric form, finding the solution would be easier. To explain how it can be converted to a symmetric form, we show for the first and second equation would be the same. First we subtract the first term from all terms of the minimizer and shift it out and repeat this for the second term. Consequently, two equal equations can be written as,

$$V_0(x) = q(x) + \alpha V_0(Ax) + \min\{0, \beta + \alpha f(x)\}, \tag{20}$$

$$V_0(x) = q(x) + \beta + \alpha V_1(Ax + b) + \min\{-(\beta + \alpha f(x)), 0\}. \tag{21}$$

By repeating for the second equation, (21) can consequently be rewritten as

$$V_0(x) = q(x) - \frac{1}{2}|\beta + \alpha f(x)| + \frac{1}{2}(\beta + \alpha V_0(Ax) + \alpha V_1(Ax + b)), \tag{22}$$

$$V_1(x) = q(x) - \frac{1}{2}|\beta - \alpha f(x)| + \frac{1}{2}(\beta + \alpha V_0(Ax) + \alpha V_1(Ax + b). \tag{23}$$

When solving this equation with the proposed $q(x)$, presenting the cost of regulation is challenging. In this case, we assume that instead of $q(x)$, another function as $\bar{c}(x, i)$ represents the first term of cost function. The, the Bellman equation for this new cost function can be written as

$$V_0(x) = \bar{c}_0(x) - \frac{1}{2}|\beta + \alpha f(x)| + \frac{1}{2}(\beta + \alpha V_0(Ax) + \alpha V_1(Ax + b), \tag{24}$$

$$V_1(x) = \bar{c}_1(x) - \frac{1}{2}|\beta - \alpha f(x)| + \frac{1}{2}(\beta + \alpha V_0(Ax) + \alpha V_1(Ax + b).. \tag{25}$$

with the goal to define $\bar{c}(x, i)$ in such a way that the right side of both equations in (25) is equal. However, after finding $\bar{c}(x, z)$, t needs to be meaningful and should be justified. To do so, it can be written as

$$\bar{c}_0(x, i) - \frac{1}{2}|\beta + \alpha f(x)| = q(x), \tag{26}$$

$$\bar{c}_1(x, i) - \frac{1}{2}|\beta - \alpha f(x)| = q(x). \tag{27}$$

By applying z, (9) can be written as a single equation

$$\bar{c}_0(x, z) = q(x) + \frac{1}{2}|\beta + (1 - 2z)\alpha f(x)|. \tag{28}$$

Now by replacing (28) in (26), the right side for both equations meaning that V0 (x) = V1 (x). Furthermore, the value function can be written without an index. The Bellman equation for the new cost function can be written as

$$V(x) = q(x) + \frac{1}{2}(\beta + \alpha V(Ax) + \alpha V(Ax + b)). \tag{29}$$

Inasmuch as the value function is only dependent on $x$, we solve the problem in this paper by proposing a quadratic solution for the value function. The proposed solution can be written as

$$V(x) = (x - \theta)^T P(x - \theta) + v, \tag{30}$$

where $\theta$ is a constant $n \times 1$ vector, $v$ is a small positive value, and $P$ is a constant $n \times n$ positive definite matrix. What remains is to determine these unknowns. To do so, the proposed solution is replaced in (30) and the equation can be written in a quadratic form of x. By setting the coefficient of the quadratic form equal to zero, it can be solved as

$$P = Q + \alpha A^T P A, \tag{31}$$

$$\theta = P^{-1}(I - \alpha A^T)^{-1}\left(Q_r - \frac{1}{2}\alpha A^T P b\right), \tag{32}$$

$$v = \frac{1}{1 - \alpha}\left(r^T Q r + \frac{1}{2}\beta + (\alpha - 1)\theta^T P \theta + \frac{1}{2}\alpha b^T P b - \alpha b^T P \theta\right). \tag{33}$$

To show the solution is reasonable, we first need to calculate f(x), which is used in the optimal policy as

$$f(x) = x^T \delta + \zeta, \tag{34}$$

where $\delta$ is a $2 \times 1$ constant vector and $\zeta$ is a constant scalar defined as

$$\delta = -2A^T P \theta, \quad \zeta = \theta^T P \theta - 2b^T P \theta. \tag{35}$$

Equation (35) shows that f(x) is the sum of weighted states and a constant, which can be rewritten as

$$\sum_{k=1}^{n}(\delta_k x_k - m_k), \quad (36)$$

where $\delta_k$ is a gain weight for each state and $m_k$ represents a specific reference. It seems that $f(x)$ is representing the sum of an error. Considering (36) shows that if this error is less than $\frac{1}{\alpha}\beta(2z-1)$, it tends to turn on the switch. However if the last state of the switch is off, this threshold is negative and the condition to turn the switch on is more difficult than the time the switch was on before and vise versa for the second equation.

IV. FEEDBACK DESIGN FOR BUCK CONVERTERS

In this paper, a buck (step-down) converter is considered as a switched-mode-controlling system. A two-state switch is used to control the output voltage of the circuit. The continuous state-space equation for a buck converter demonstrated in "Fig. 1" can be written as

$$\dot{x}(t) = A_c x(t) + b_c u(t), \quad (37)$$

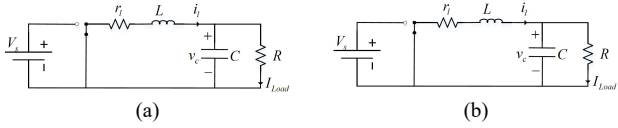

(a)             (b)

Figure 1. (a) conductive mode $u(t) = 1$ and (b) The switch is connected to the ground

where $x(t)$ is a $2 \times 1$ vector representing the state variables of the buck converter, $A_c$ is a constant $2 \times 2$ matrix, $b_c$ is a constant $2 \times 1$ vector, and $u(t)$ presents the position of the switch at the time t, and all can be defined as

$$\dot{x}(t) = \begin{bmatrix} v_c(t) \\ i_l(t) \end{bmatrix}, \quad u(t) \in \{0,1\}, \quad (38)$$

where $v_c(t)$ represents the capacitor voltage and $i_l(t)$ represents the inductor current. In this paper, $u(t) = 1$ means that the switch is connected to the power source, and $u(t) = 0$ means that the switch is connected to the ground of the circuit. Moreover, $A_c$ and $b_c$ can be written as

$$A_c = \begin{bmatrix} -\frac{1}{RC} & \frac{1}{C} \\ -\frac{1}{L} & \frac{r_l}{L} \end{bmatrix}, \quad b_c = \begin{bmatrix} 0 \\ \frac{V_s}{L} \end{bmatrix} \quad (39)$$

where $L$ represents the inductance and $r_l$ accounts for its losses, and $C$ denotes the capacitances representing the input source voltage, and $R$ represents the output load resistor [30].

Considering the proposed optimal control model is in the discrete region, the state-space model needs to be converted into the same region. In order to do that, a continuous model can be converted to a discrete one by the transformation of

$$A = e_c^A T, \quad (40)$$

$$b = \int_0^\lambda e_c^A T d\lambda b_c, \quad (41)$$

where $T$ is the sample time. To simulate the buck converter and apply the proposed optimal controller, we normalized all the parameters. In this case, the value of the load resistor $R$, switching step time, and the input voltage source $Vs$ are considered base values. The applied parameter is shown in "Table I".

In the simulation, we try to show that the proposed controller works well in the start time and in the load changes, which represents the transient period. The steady state results are also analyzed.

As "Fig. 2" shows, the buck converter reaches steady state after 8 msec with a reasonable overshoot. When the voltage is less than the reference voltage, the switch stays in the "on" position and then responds like a pulse width modulation (PWM) signal.

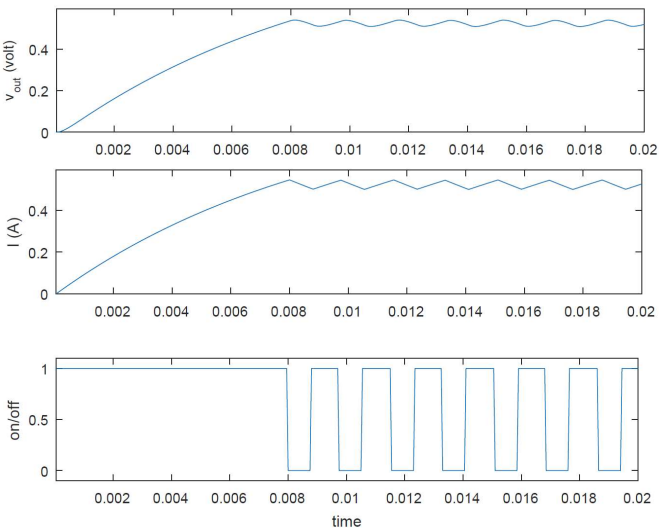

Figure. 2 Simulation results for startup from the initial condition zero

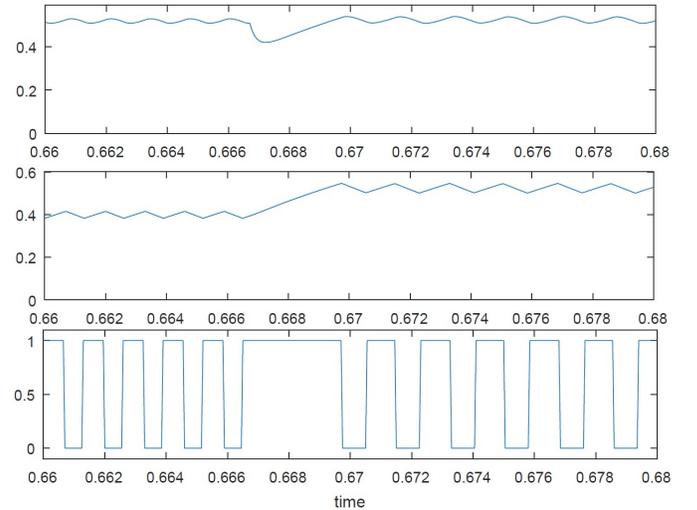

Figure. 3 Simulation results for the load increase from nominal to 1.3 nominal

In "Fig. 3 and 4", the controller response to a load change is shown. In Fig. 3, the load increases for %30, and in Fig. 4 it decreases to its nominal value. Another advantage of this model is that when the load changes, the frequency, slightly changes.

This is the dominance of not being restricted to the PWM frequency, and since the changes in frequency are negligible, it does not affect the electromagnetic interference (EMI) filter design.

In addition, "Fig. 5" demonstrates that in steady states, the system responds well. It goes without saying that by increasing the switching penalty factor $\beta$, the number of switching

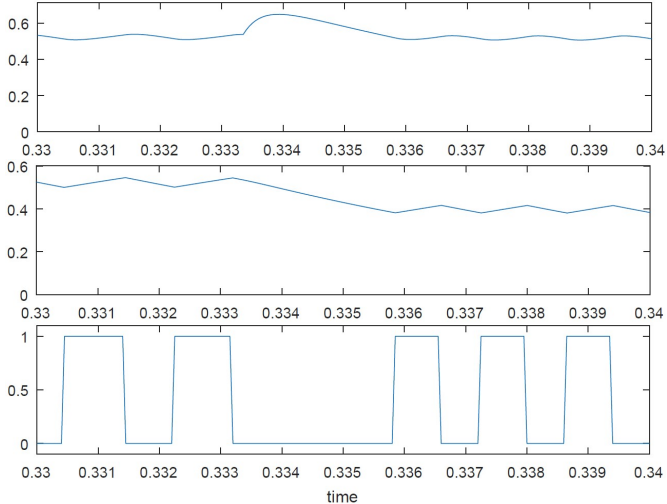

Figure. 4 Simulation results for the load decrease for going back from 1.3 nominal to the nominal value

decreases and results in a less accurate output and a bigger voltage ripple.

The other important aspect of this controller is how well it responds to turbulence which was tested by adding random noise with a maximum magnitude of 0:1 to the input. The results are shown in "Fig. 6".

## V. CONCLUSION

In this paper, a new uniform model was presented to model a DC-DC unfixed-frequency switch-mode step-down converter. The constrained optimal controller was studied, and an approximate solution to the associated Bellman equation has been proposed. The proposed modelling and controller have tackled the drawbacks of traditional PI-based controllers. Additionally, the proposed method is feasible for the transient time, functions well in the steady state. The proposed model and controller have been implemented in MATLAB to validate their effectiveness. As shown with the simulation results, the proposed controller behaves well in different scenarios, including start up, load change, and steady state.

Table I. Converter and controller parameters

| Converter Parameter | | | Controller Parameter | |
|---|---|---|---|---|
| parameter | In S.I | In p.u | parameter | value |
| $L$ | 1 mH | 27.9 | $\alpha$ | 0.9999 |
| $C$ | 220 uF | 497 | $\beta$ | 10 |
| $r_l$ | 1.5 Ω | 0.17 | $f_s$ | 20 kHz |
| $R$ | 9 Ω | 1 | | |
| $V_s$ | 20 V | 1 | $Q$ | $\begin{bmatrix} 1 & 0 \\ 0 & 0 \end{bmatrix}$ |
| $V_r$ | 8 V | 0.4 | | |

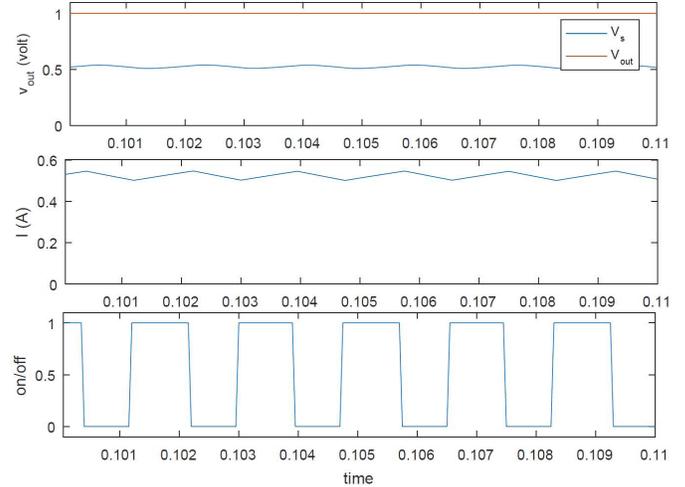

Figure. 5 Simulation results for the steady state

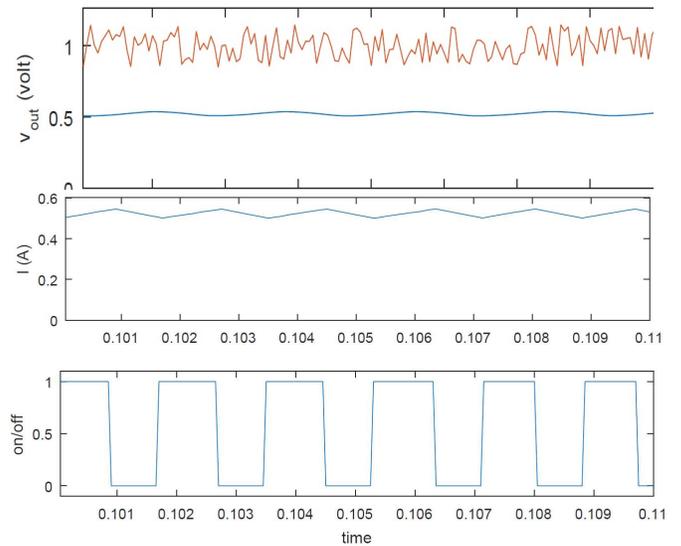

Figure. 6 Simulation results for a random noise with amplitude of 0.3 is applied to the input voltage